\newcommand{\version}{March 2, 2022}
\documentclass[letterpaper,oneside,11pt,pdftex]{article}
\pdfoutput=1
\usepackage[bindingoffset=0.0cm,textheight=22.60cm,hdivide={*,16.25cm,*}, vdivide={*,22.60cm,*}]{geometry}
\usepackage{amsmath,amsfonts,amssymb}
\usepackage{mathtools}
\usepackage[pdftex]{graphicx}
\usepackage[margin=1cm,font=small]{caption}
\usepackage[T1]{fontenc}
\usepackage[utf8]{inputenc}
\usepackage{fouriernc}
\newcommand{\id}{\mathbb{1}}

\usepackage[numbers,square,comma,sort&compress]{natbib}
\usepackage[pdftex,hyperref,svgnames]{xcolor}
\usepackage[pdftex,bookmarksnumbered=true,breaklinks=true,%
colorlinks=true,linktocpage=true,linkcolor=MediumBlue,citecolor=ForestGreen,urlcolor=DarkRed]{hyperref}

\makeatletter
\AtBeginDocument{
  \hypersetup{
    pdfkeywords = {\keywords},
    pdftitle = {\@title},
    pdfauthor = {\@author}
  }
}
\makeatother

\makeatletter

 \@addtoreset{equation}{section}
 \makeatother

\renewcommand{\d}{\delta}

\newcommand{\vth}{\vartheta}

\newcommand{\sgn}[1]{\textrm{sgn}\!\left(#1\right)}
\newcommand{\nn}{\nonumber}
\newcommand{\eqnref}[1]{Eq. \eqref{#1}}

\newcommand{\ct}{c_{\textrm{T}}}

\newcommand{\vc}{v_{\textrm{c}}} %% critical velocity

\newcommand*{\mat}[1]{\mathrm{#1}}

\renewcommand{\Re}{\mathrm{Re}}

\newcommand{\txt}[1]{\textrm{#1}}

\DeclarePairedDelimiter\abs{\lvert}{\rvert}
\newcommand{\coleq}{\vcentcolon=}
\title{\texorpdfstring{\begin{flushright}
        {\small LA-UR-21-26126}
       \end{flushright}\vspace{2em}}{}%
       How to determine limiting velocities of dislocations in anisotropic crystals}

\author{Daniel N. Blaschke}

\date{\version}

\newcommand{\keywords}{dislocations in crystals, dislocation mobility, crystal plasticity}

\graphicspath{{./}{./figures/}}

\begin{document}

 \maketitle

 \thispagestyle{empty}
 \begin{center}
 \vspace{-0.3cm}
 Los Alamos National Laboratory, Los Alamos, NM, 87545, USA
 \\[0.5cm]
 \ttfamily{E-mail: dblaschke@lanl.gov}
 \end{center}

\begin{abstract}
In the continuum limit, the theory of dislocations in crystals predicts a divergence in the elastic energy of the host material at a crystal geometry dependent limiting (or critical) velocity $\vc$.
Explicit expressions for  $\vc$ are scattered throughout the literature and are available in analytic form only for special cases with a high degree of symmetry.
The fact that in some cases (like pure edge dislocations in fcc) $\vc$ happens to coincide with the lowest shear wave speed of a sound wave traveling parallel to the dislocation's gliding direction has led to further confusion in the more recent literature.
The aim of this short review therefore is to provide a concise overview of the limiting velocities for dislocations of arbitrary character in general anisotropic crystals, and how to efficiently compute them, either analytically or numerically.
\end{abstract}

% \vspace{1cm}
% \newpage
%\tableofcontents
% \newpage

\section{Introduction and background}
\label{sec:intro}
%%%%%%%%%%%%%%%%%%%%%%%%%%%%%%%%%%%%%%%%%%%%%%%%%%%%%%%%%

High rate plastic deformation in metals and other crystalline materials is governed by the mobility of dislocations, as it determines the glide time between obstacles (grain boundaries, impurities, other defects, etc.), thereby affecting Orowan's relation \cite{Hansen:2013,Luscher:2016,Blaschke:2019a,Blaschke:2021impact}.
In particular, the highest achievable strain rates are determined not only by the mobile dislocation density but also by the limiting velocities of dislocations.
These can be calculated from linear elasticity theory in the limit of perfect, steady-state dislocations, and neglecting details of the dislocation core \cite{Teutonico:1961,Teutonico:1962,Blaschke:2017lten,Blaschke:2020MD}.
At these limiting velocities, $\vc(\vartheta)$, which in general depend on dislocation character angle $\vartheta$ and which in the isotropic limit coincide with the transverse sound speed, the dislocation self energy is predicted to diverge \cite{Weertman:1961,Weertman:1980,Blaschke:2017lten}.
Likewise, the dislocation drag coefficient, which accounts for dislocation motion being impeded by phonon scattering, is also predicted to diverge \cite{Blaschke:2018anis} --- even for accelerating dislocations \cite{Blaschke:2020acc}.
Nonetheless, the $\vc(\vartheta)$ need not necessarily be viewed as hard barriers, since it was shown that these velocities can in principle be overcome when the dislocation core is taken into account in a regularising fashion, see \cite{Markenscoff:2008,Pellegrini:2018,Pellegrini:2020}.
Thus within real crystals, the $\vc$ can be seen as dislocation velocities that, at the very least, are hard to overcome.
Indeed, a number of Molecular Dynamics (MD) simulations~\cite{Olmsted:2005,Marian:2006,Daphalapurkar:2014,Tsuzuki:2008,Tsuzuki:2009,Oren:2017,Ruestes:2015,Gumbsch:1999,Li:2002,Jin:2008,Peng:2019} as well as some experimental data~\cite{Nosenko:2007} suggest that
dislocations can reach transonic or even supersonic velocities in certain cases.
Most strikingly, pure edge dislocations in some fcc metals seem to asymptotically approach the lowest shear wave speed (which in this special case happens to coincide with $\vc$) until some critical stress above which transonic motion becomes possible \cite{Olmsted:2005,Marian:2006,Daphalapurkar:2014,Tsuzuki:2008,Tsuzuki:2009,Oren:2017}.

Even though the theoretical tools to compute the limiting velocities of dislocations were derived many decades ago \cite{Teutonico:1961,Weertman:1962,VanHull:1962,Teutonico:1962}, most of the modern literature assumes (ad hoc) that $\vc$ coincides with the lowest shear wave speed $c_s$ of sound waves traveling in the dislocation glide direction.
This is not the case in general, though it is possible that the two velocities coincide in some cases.
None of Refs. \cite{Olmsted:2005,Marian:2006,Daphalapurkar:2014,Tsuzuki:2008,Tsuzuki:2009,Oren:2017,Ruestes:2015,Gumbsch:1999,Li:2002,Jin:2008,Peng:2019,Krasnikov:2010,Barton:2011,Lloyd:2014JMPS,Luscher:2016,Gurrutxaga:2020}  make any reference to this subtlety, and this is the main reason to write this short review.

In fact, many of the MD simulations cited above were concerned with fcc edge dislocations where indeed $\vc=c_s$, see \cite{Olmsted:2005,Marian:2006,Daphalapurkar:2014,Tsuzuki:2008,Tsuzuki:2009,Oren:2017}.
Others, like Refs. \cite{Gumbsch:1999,Li:2002,Jin:2008}, studied bcc tungsten whose second order elastic constants are ``almost isotropic'' in the sense that $c_{11} \approx c_{12}+2c_{44}$ so that $\vc\approx\ct\approx c_s$.
Some authors, however, studied pure screw dislocations in fcc Cu \cite{Oren:2017} and Ni \cite{Olmsted:2005,Marian:2006} at room temperature and misinterpreted their results to report supersonic dislocation motion by overlooking that for fcc screw dislocations $\vc>c_s$; this was recently clarified in Ref. \cite{Blaschke:2020MD}.

Here, we aim to review how to compute limiting dislocation velocities in general and for all crystal geometries and slip systems, to present analytical expressions in cases where they are readily available, and simple ``recipes'' to compute them efficiently otherwise.
Our main objectives are clarity and brevity so that these results can be easily incorporated into future MD simulation (and other) studies.
In so doing, we also make some points that (to my knowledge) have not been elucidated in the literature before, namely the explicit analytic expression of \eqref{eq:vcedge_special2} and the subtle cancellations that can occur for slip systems with reflection symmetry (in addition to fcc) discussed after \eqnref{eq:detnn}.

\section{Limiting velocities of dislocations: Special cases}
\label{sec:specialcases}
The displacement field $u_i$ of any dislocation must fulfill the following differential equations:
\begin{align}
\partial_i \sigma_{ij}  &= \rho \ddot{u}_j
\,, &
\sigma_{ij} &= C'_{ijkl} \partial_l u_{k}
\label{eq:diffeqns1}
\end{align}
in coordinates aligned with the dislocations, i.e. $\hat z$ is aligned with the dislocation line and $\hat y$ is parallel to the slip plane normal.
The components of the tensor of second order elastic constants (SOEC) is always measured in Cartesian coordinates that are aligned with the crystal axes, and thus this tensor must be rotated into the present coordinate basis, i.e.:
\begin{align}
	C'_{ijkl} = U_{ii'}U_{jj'}U_{kk'}U_{ll'}C_{i'j'k'l'}
\end{align}
with rotation matrix $U$.
If Cartesian unit vector $\hat t$ is parallel to the dislocation line, and $\hat n_0$ is the slip plane normal in Cartesian coordinates aligned with the crystal, and a third unit vector $\hat m_0=\hat n_0\times \hat t$ is normal to the former two, then the rotation matrix is easily determined by stacking the three row vectors into a matrix, i.e.:
\begin{align}
U&=\left(\begin{array}{c}
 \hat m_0^T \\ \hat n_0^T \\ \hat t^T
\end{array}\right)
\,,
\end{align}
so that $U\cdot \hat m_0=\hat x$, $U\cdot \hat n_0=\hat y$, and $U\cdot \hat t=\hat z$.
Note that the line direction depends on the dislocation character angle $\vartheta$ because both slip plane normal $\hat n_0$ and Burgers vector $\vec{b}$ are determined by the slip system under consideration and by definition $\vartheta$ is the angle between $\hat t$ and $\vec{b}$.
Thus, $\hat t(\vartheta) =\hat b\cos\vartheta+\hat b\times\hat n_0\sin\vartheta$ with unit vector $\hat b=\vec{b}/\abs{b}$.
Consequently, also $\hat m_0$ and $U$ are character dependent.

Assuming the dislocation moves at constant velocity $v$ in the direction perpendicular to its line sense and the slip plane normal, i.e. parallel to $\hat x$, the differential equations \eqref{eq:diffeqns1} simplify further, since the dislocation can then only depend on time via the combination $x'\equiv x-vt$.
Thus, a time derivative is related via the chain rule to a spatial derivative in $x$ and hence \cite{Saenz:1953}:
\begin{align}
C'_{ijkl} \partial_i \partial_l u_{k}  &= -v\rho\partial_{x}^2{u}_j
\,. \label{eq:diffeqns2}
\end{align}
In Ref. \cite{Bullough:1954} Bullough and Bilby make the ansatz
\begin{align}
u_j(x',y) &= \sum_{n} C_n P_{jn}\exp\left(-s\lambda_ny+isx'\right)
\,, \label{eq:Bulloughansatz}
\end{align}
where the $P_{jn}$ are functions of the elastic constants and the dislocation velocity, $C_n$ are arbitrary complex constants (determined by the boundary conditions for dislocations), and $\sgn{y}\Re(\lambda_n)>0$ ensures that the displacement vanishes at $\abs{y}\to\infty$.
We will not discuss this solution in full detail here: the point we would like to make presently is that the $\lambda_n$ are functions of the dislocation velocity and whenever $\Re(\lambda_n(v))\to0$, that velocity is a limiting velocity because the elastic self energy of the dislocation will diverge there.
In general, upon inserting the ansatz \eqref{eq:Bulloughansatz} into the differential equations \eqref{eq:diffeqns2}, the $\lambda_n$ follow from solving the determinantal sextic equation \cite{Teutonico:1961,Teutonico:1962}
\begin{align}
\det\left|C'_{i2k2}\lambda^2-i\lambda\left(C'_{i1k2}+C'_{i2k1}\right)-C'_{i1k1}+\rho v^2\delta_{ik}\right|=0
\,. \label{eq:sextic}
\end{align}
The solutions $\lambda_n$ are complex in general, and for each of the two half-planes $\abs{y}\neq0$, the ones with $\sgn{y}\Re(\lambda_n)>0$ are considered in the sum over $n$ within \eqref{eq:Bulloughansatz}.

In order to study pure screw or pure edge dislocations, the rotated tensor of SOEC must fulfill the following symmetry requirements, shown here in Voigt notation which maps index pairs to single digits, $(11, 22, 33, 32/23, 31/13, 21/12) \rightarrow (1, 2, 3, 4, 5, 6)$:
\begin{align}
	C'_{\alpha\beta} = \left(\begin{array}{cccccc}
		c'_{11} & c'_{12} & c'_{13} & 0 & 0 & c'_{16} \\
		c'_{12} & c'_{22} & c'_{23} & 0 & 0 & c'_{26} \\
		c'_{13} & c'_{23} & c'_{33} & c'_{34} & c'_{35} & c'_{36} \\
		0 & 0 & c'_{34} & c'_{44} & c'_{45} & 0 \\
		0 & 0 & c'_{35} & c'_{45} & c'_{55} & 0 \\
		c'_{16} & c'_{26} & c'_{36} & 0 & 0 & c'_{66}
	\end{array}
	\right)
	,\label{eq:SOECreflectionplane}
\end{align}
i.e. the six components $c'_{14}$, $c'_{15}$, $c'_{24}$, $c'_{25}$, $c'_{46}$, and $c'_{56}$ must vanish, see Refs. \cite{Foreman:1955} and \cite[Sec. 13-4]{Hirth:1982}.
This ensures that $u_3=0$ implies $\partial_i\sigma_{i3}=0$ and likewise that $u_1=0=u_2$ implies $\partial_i \sigma_{i1}=0=\partial_i \sigma_{i2}$.
Note that in the present coordinates, displacement field $u_i$ for a straight dislocation can only depend on $x$, $y$, and $t$, but not on $z$.
This latter property implies that non-vanishing components $c'_{34}$ and $c'_{35}$ are allowed since they do not enter the differential equations above for pure screw or pure edge dislocations.
On the other hand, the stronger condition $c'_{34}=0=c'_{35}$ implies that the $x_1$, $x_2$ plane is a reflection plane (and then $\sigma_{33}=0$ for pure screw dislocations rather than the weaker $\partial_3\sigma_{33}=0$).
If these requirements are fulfilled, the sextic \eqnref{eq:sextic} decouples into a quadratic equation for pure screw dislocations and a quartic equation for pure edge dislocations.
We review solutions to these two simpler equations in the following two subsections.
Note, however, that since the dislocation line direction (w.r.t. to the crystal axes) is dislocation character dependent (because the Burgers vector direction is the same for all character angles of a given slip system), the rotated tensor of SOEC, $C'_{\alpha\beta}$, is \emph{different} for pure screw than pure edge dislocations within the same slip system.

\subsection{Pure screw dislocations with reflection symmetry}
%%%%%%%%%%%%%%%%%%%%%

If symmetry property \eqref{eq:SOECreflectionplane} is fulfilled, the case of a pure screw dislocation follows from setting index $i=3=k$ within \eqnref{eq:sextic}.
This leads to the simple quadratic equation \cite{Bullough:1954,Teutonico:1961}
\begin{align}
c'_{44}\lambda^2-2ic'_{45}\lambda - c'_{55} + \rho v^2=0
\,,
\end{align}
with the two solutions
\begin{align}
\lambda_n &= i\frac{c'_{45}}{c'_{44}} \pm \frac{1}{c'_{44}} \sqrt{{c'_{44}}{c'_{55}}  - {(c'_{45})^2} - {c'_{44}}{\rho v^2}}
\,,
\end{align}
whose real part vanishes at the limiting velocity \cite{Bullough:1954,Teutonico:1961,Markenscoff:1984JE,Blaschke:2020MD}
\begin{align}
v_c^\text{screw} &= \sqrt{\frac{1}{\rho}\left(c'_{55}-\frac{(c'_{45})^2}{c'_{44}}\right)}
\,. \label{eq:vcscrew}
\end{align}
Examples of pure screw dislocations with reflection symmetry in cubic and hexagonal crystals include
all 12 fcc slip systems, and the hcp slip systems with Burgers vectors $b=\langle2,1,1,0\rangle$ including basal, prismatic, and pyramidal slip planes.
Of course, other crystal geometries (such as tetragonal, orthorhombic, and others) can also exhibit slip systems with a reflection symmetry for pure screw dislocations. 
In the isotropic limit, $C_{ijkl}$ is invariant under rotations, and $c'_{45}=0$ and $c'_{55}$ equals shear modulus $\mu$ so that $\vc$ is simply given by the transverse sound speed $\ct$.

\subsection{Pure edge dislocations with reflection symmetry}
\label{sec:edgereflection}

For pure edge dislocations, we need to study the 2x2 matrix of \eqnref{eq:sextic} with $i=1,2$ and $k=1,2$ (provided symmetry property \eqref{eq:SOECreflectionplane} is fulfilled).
The determinant in this case reads \cite{Teutonico:1961}
\begin{align}
	\left(c'_{66}  \lambda^2 - 2ic'_{16}\lambda - c'_{11} + \rho v^2\right)
	\left(c'_{22}\lambda^2 - 2ic'_{26}\lambda - c'_{66} + \rho v^2\right)
	- \left(c'_{26}\lambda^2 - i\lambda\left(c'_{12}+c'_{66}\right) - c'_{16}\right)^2=0
	\,. \label{eq:quarticeq_vc}
\end{align}
In general, this quartic equation has four roots of the form $\lambda_n = \pm p_n + i q_n$.
The analytic expressions can be determined explicitly, but are very tedious and lengthy.
Thus, it is in practice more efficient to first plug in numerical values for the elastic constants in order to determine $\lambda$ as a function of velocity, and to then numerically determine the limiting velocity by setting the product of the real parts of all four roots $\lambda_n$ to zero, i.e. by numerically solving $\prod_{n=1}^4 p_n(v_c)=0$.
This is the strategy employed in the open source code PyDislocDyn \cite{pydislocdyn} developed by the present  author, which can be used to not only calculate this special case, but also any limiting velocity for steady state dislocations of arbitrary character angle in arbitrary crystal and slip system geometry.

Examples of pure edge dislocations with reflection symmetry in cubic and hexagonal crystals include  hcp basal and prismatic slip systems with Burgers vectors $b=\langle2,1,1,0\rangle$, as well as the \{112\} slip planes in bcc crystals.
The latter bcc slip systems have $c'_{26}\neq0$ and need to be treated the way we have just discussed above.
The basal and prismatic hcp slip systems on the other hand feature $c'_{16}=0=c'_{26}$ which leads to additional simplifications which we discuss next.

\subsubsection*{The special case of $\mathbf{c'_{16}=0=c'_{26}}$}

If $c'_{16}=0=c'_{26}$, \eqnref{eq:quarticeq_vc} simplifies to the following quadratic equation in $\lambda^2$:
\begin{align}
	&	c'_{22}c'_{66} \left(\lambda^4 - q\lambda^2 + s\right)
	=0
	\,, \qquad\text{with}
	\nn\\
	&q= \frac{1}{c'_{22}c'_{66}}\left[c'_{22}\left(c'_{11} - \rho v^2\right)+c'_{66}\left( c'_{66} - \rho v^2\right)-\left(c'_{12}+c'_{66}\right)^2\right]
%	\nn\\
%	&\phantom{q}= \frac{1}{c'_{22}c'_{66}}\left[c'_{22}c'_{11} - (c'_{12})^2 - 2c'_{12}c'_{66} -  \left(c'_{22}+c'_{66}\right) \rho v^2\right]
	\,,\nn\\
	&s=\frac{1}{c'_{22}c'_{66}} \left(c'_{11} - \rho v^2\right)\left( c'_{66} - \rho v^2\right)
	\,. \label{eq:quadraticeq_vc}
\end{align}
There are two distinct cases to consider \cite{Teutonico:1961} (see also \cite{Markenscoff:1984}):
\begin{enumerate}
\item If $q\ge0$, the limiting velocity is\footnote{
A further special case which falls into this category regarding its limiting velocity, $c'_{12}+c'_{66}=0$, is considered in Ref. \cite{Markenscoff:1985a}, though it is unclear if any slip systems fulfill this condition:
None of the cubic, hexagonal, tetragonal, and orthorhombic slip systems checked by the present author exhibit this special property.
}
\begin{align}
\vc^\txt{edge}=\sqrt{\frac{\min(c'_{11},c'_{66})}{\rho}}
\,, \label{eq:vcedge_special1}
\end{align}
because then one of the two solutions to $\lambda_n^2$ tends to zero.
More precisely, the condition for this expression to be a limiting velocity for pure edge dislocations is $q(\vc)\ge0$.
Many hcp basal and prismatic slip systems fall into this category.
In the isotropic limit, $c'_{66}=\mu$ and $\vc=\ct$, as expected.

\item If on the other hand $q<0$, then the square root in the solution to $\lambda^2$ would become imaginary as $\rho v^2\to \min(c'_{11},c'_{66})$, and in this case there exists a smaller velocity that renders the real part of at least one solution to $\lambda_n$ zero; in particular $v_c$ is determined in this case from the non-linear equation \cite{Teutonico:1961}
\begin{align}
q&=-2\sqrt{s}
\,.
\end{align}
Its solution can be derived analytically and thus the smallest (positive) limiting velocity derived in this way reads:
\begin{align}
\vc^\txt{edge} &= \frac{\sqrt{2A - B}}{\sqrt{\rho}\left|c'_{22} - c'_{66}\right|}
\,, \qquad\text{with} \nn\\
A &= (c'_{12} + c'_{66})\sqrt{c'_{22}c'_{66}\left[c'_{11}(c'_{66} -c'_{22} )+ (c'_{12})^2 + c'_{66}(2c'_{12} + c'_{22}) \right]}
\,,\nn\\
B &= \left[ c'_{11}c'_{22}(c'_{66}-c'_{22}) + (c'_{12})^2(c'_{22} + c'_{66}) + 2c'_{12}c'_{22}c'_{66}  + 2(c'_{66})^2(c'_{12}+c'_{22})\right]
\,. \label{eq:vcedge_special2}
\end{align}
The elastic stability criterion $c'_{11}c'_{22}>(c'_{12})^2$ (see \cite{Teutonico:1961}) together with $q<0$ ensures that this expression is real.
An example for this case is the basal slip system of Zn.
\end{enumerate}

\section{Limiting velocities of dislocations: The most general case}
\label{sec:generalcase}
%%%%%%%%%%%%%%%%%%%%%%%%%%%%%%%%

The most general case can in principle be solved in the same way by studying the full sextic equation \eqref{eq:sextic}.
This has been done by Teutonico in Ref. \cite{Teutonico:1962}, but the determination of explicit numerical values for the limiting velocities is tedious and numerically not very efficient.
A computationally faster and thus better way to study the limiting velocities in this case is to employ the so-called integral formalism which is based on the work of Stroh and others \cite{Stroh:1962,Barnett:1973,Asaro:1973}, see \cite{Bacon:1980} for a review as well as \cite[pp.~467--478]{Hirth:1982}.
The main ideas leading to this solution are summarized as follows \cite{Blaschke:2017lten}.
Stroh made an ansatz for a solution in Cartesian crystal coordinates which depends on perpendicular unit vectors, $\vec{m}$ and $\vec{n}$ which are normal to the dislocation sense vector $\vec{t}$.
Thus, the differential equation is converted to an eigenvalue problem which, due to Voigt symmetry, can be formulated in terms of a 6-dimensional vector and associated $6\times6$ matrix $\mat{N}$, known as the ``sextic formalism''.
Since unit vectors $\vec{m}$ and $\vec{n}$ within this ansatz are only defined up to an arbitrary angle $\phi$, Barnett, Lothe, and others \cite{Barnett:1973,Asaro:1973} realized that the solution can be written in terms of the average matrix
\begin{align}
\langle \mat N\rangle&=\frac1{2\pi}\int_0^{2\pi}\mat N d\phi
=\begin{pmatrix}
\mat S & \mat Q \\
\mat K & \mat S^T
\end{pmatrix}
\,,
\end{align}
with
\begin{align}
\mat S&=-\frac1{2\pi}\int_0^{2\pi}(nn)^{-1}(nm)\,d\phi  \,, &
\mat S^T&=-\frac1{2\pi}\int_0^{2\pi}(mn)(nn)^{-1}d\phi  \,,\nn\\
\mat Q&=-\frac1{2\pi}\int_0^{2\pi}(nn)^{-1}d\phi  \,, &
\mat K&=-\frac1{2\pi}\int_0^{2\pi}\left[(mn)(nn)^{-1}(nm)-(mm)\right]d\phi
\,, \label{eq:thematrix}
\end{align}
where we have employed the shorthand notation $(ab)_{jk}\coleq a_i \left(C_{ijkl}-\rho v_iv_l \d_{jk}\right) b_l$ and unit vectors
\begin{align}
\vec{m}(\vth,\phi)&=\hat{m}_0(\vth)\cos(\phi) + \hat{n}_0\sin(\phi)
\,,\nn\\
\vec{n}(\vth,\phi)&=\hat{n}_0\cos(\phi) - \hat{m}_0(\vth)\sin(\phi)
\,,\label{eq:vecs-sol}
\end{align}
depend not only on polar angle $\phi$, but also on the dislocation character angle $\vartheta$.
In particular, if $\hat{n}_0$ is the slip plane normal and $\hat{t}(\vth)$ is the sense vector of the dislocation, then $\hat{m}_0(\vth) = \hat{n}_0 \times \vec{t}(\vth)$.
It is also easy to work out that all integrands in \eqref{eq:thematrix} are $\pi$-periodic in $\phi$ due to \eqref{eq:vecs-sol}, so that $\int_0^{2\pi}(\ldots)\, d\phi=2\int_0^{\pi}(\ldots) \,d\phi$.
The beauty of this formalism lies in the fact that the computation of the eigenvalues can be completely circumvented because the eigenvectors happen to be independent of the choice of basis $\vec{m}$, $\vec{n}$ which is parametrized by polar angle $\phi$.
This is of course due to the symmetry properties of the steady-state problem.

Within this `integral method', whose derivation is very nicely explained in the review article of Ref. \cite{Bacon:1980}, the solution to the dislocation displacement gradient field takes the form $\partial_k u_{j}(r,\phi)={\tilde{u}_{jk}(\phi)}/{r}$, where
\begin{align}
\tilde{u}_{jk}(\phi)&=\frac{-b_l}{2\pi}\left\{n_k\left[(nn)^{-1}(nm) S\right]_{jl} - m_k S_{jl} + n_k(nn)^{-1}_{ji}K_{il}\right\}\
\label{eq:ukl-sol}
\end{align}
is a function of the Burgers vector $\vec{b}$, and polar angle $\phi$.
Furthermore, matrices $\mat S$ and $\mat K$ are independent of position and thus need to be computed only once.
Due to the explicit character angle $\vartheta$ dependence, this solution can be easily applied to any dislocation of mixed character.

Having this solution at hand, it is no longer necessary to solve a sextic equation in order to determine the limiting velocities.
Instead, one has to study the determinant of a $3\times3$ matrix, namely \cite{Blaschke:2017lten,Bacon:1980}
\begin{align}
0 &= \det(nn) = \det\left(\vec{n}\cdot\mat{C}\cdot\vec{n}-\rho \left(\vec{n}\cdot\vec{v}\right)^2\id\right)
\,, \label{eq:detnn}
\end{align}
where $\rho$ is the material density, $\mat{C}$ is once more the tensor of second order elastic constants, the 3x3 identity matrix is denoted as $\id$,
%and $\vec{v}$ is the dislocation velocity vector which lies in the glide plane spanned by $\vec{t}$ and $\hat{m}_0$.
and $\vec{v}=v\hat{m}_0$ is the dislocation velocity vector.
For any given dislocation character angle $\vartheta$ there are three solutions to \eqnref{eq:detnn}, $v_n(\phi)$, and each one is a function of polar angle $\phi$ because basis vector $\vec{n}$ depends on $\phi$ according to \eqref{eq:vecs-sol}.
The smallest value for each branch $v_{c,n}=\min(v_n(\phi))$ is found by minimizing these functions in the interval $\phi\in[0,2\pi]$.
There is, however, another caveat to this problem:
A divergence at $\vc$ occurs in $\partial_k u_{j}$ because $\det(nn)=0$ implies a divergence in the inverse of matrix $(nn)$, but it is nonetheless possible that subtle cancellations in the matrix products within $\tilde{u}_{jk}$ lead to a finite result despite the vanishing determinant.
The most prominent example is the pure screw dislocation in an fcc crystal, where it was shown in Ref. \cite{Blaschke:2020MD} that such a cancellation occurs for the smallest $v_{c,n}$ computed from $\det(nn)=0$, and that the true limiting velocity is given by the \emph{second} branch of that solution instead of the first.
The true limiting velocity then of course coincides with \eqref{eq:vcscrew}.
In fact, this type of cancellation within $\tilde{u}_{jk}$ at $\det(nn)=0$ requires a high degree of symmetry within the crystal, which we identify as the reflection symmetry discussed above in Sec. \ref{sec:specialcases}.
It can be checked by direct computation, that only a subset of pure screw or edge dislocations with reflection symmetry feature this type of subtlety.

To sum up:
In order to determine the limiting velocity for a pure screw or edge dislocation with reflection symmetry, one best employs the results discussed above in Sec. \ref{sec:specialcases}.
For all other cases, including any mixed dislocation with arbitrary dislocation character angle $\vartheta$, the smallest value of $v_{c,n}$ determined from $\det(nn)=0$ after minimization with respect to polar angle $\phi$ will be the limiting velocity.
This strategy is for instance implemented in the open source code PyDislocDyn \cite{pydislocdyn} developed by the present author.

\subsection{Relation to sound speeds and the Rayleigh wave speed}
%%%%%%%%%%%%%%%%%%%%%%%%%%%%%%%%%

%A further remark is in order:
The sound speeds in the direction of dislocation motion, $\hat{v}=\vec{v}/v$ (parallel to the dislocation glide direction), are in general determined from \cite{Duff:1960,Bacon:1980}
\begin{align}
	\det\left(\hat{v}\cdot\mat{C}\cdot\hat{v}-\rho v^2\id\right)\Big|_{v=v_{\txt{shear}}}=0
	\,. \label{eq:sound1}
\end{align}
This equation clearly \emph{differs} from \eqref{eq:detnn} for the limiting dislocation velocities above, and therefore the smallest of the three solutions for sound speed $v$ in the direction of $\hat{v}$, may or may not coincide with the limiting velocity of a dislocation \cite{Blaschke:2017lten,Blaschke:2020MD}.
In fact, circling back to the example of an fcc crystal, the limiting velocity for a pure edge dislocation does happen to coincide with the lowest shear wave speed for a sound wave propagating in the same direction, but the limiting velocity of a pure screw dislocation is \emph{higher} than the corresponding lowest shear wave speed \cite{Blaschke:2020MD}.
On the other hand, all limiting velocities for dislocations must lie within the range of all shear wave speeds of the crystal, e.g. for a cubic crystal $\rho\vc^2\in\left[\min(c_{44},c'),\max(c_{44},c')\right]$ with $c'=(c_{11}-c_{12})/2$ for all dislocations and slip systems.
In the isotropic limit, $c_{44}=\mu=c'$ and there is only one shear wave speed $\ct$ which coincides with the limiting velocity regardless of the dislocation character angle.

%Before we close this subsection, a final remark is in order:
There is another velocity that plays a crucial role in dislocation dynamics, namely the generalized Rayleigh wave speed.
In a general anisotropic setting, it can take any value between 0 and the limiting velocity (depending on the elastic constants) \cite{Teutonico:1962}, whereas in the isotropic limit it always lies within the interval [$0.69\ct,0.96\ct$] (depending on Poisson's ratio $\nu$ and with typical values around $0.93\ct$ for $\nu\sim1/3$) \cite{Teutonico:1961}.
Above this velocity, the force between edge dislocations is known to change sign.
This was first pointed out in the isotropic case by Weertman \cite{Weertman:1961,Weertman:1980}.
As Weertman also pointed out, the dislocation self energy is regular at this speed and therefore the Rayleigh velocity is not a limiting velocity itself.
In general, the Rayleigh velocity can be calculated on a per character basis from the necessary and sufficient condition \cite{Barnett:1973b}
\begin{align}
	\frac{1}{2}\left(K_{11}+K_{22}\right) + \sqrt{\frac14\left(K_{11}-K_{22}\right)^2 + \left(K_{12}\right)^2}&=0
	\,, \label{eq:rayleigh}
\end{align}
with $K_{ij}(v,\vartheta)$ given in \eqnref{eq:thematrix}.
It was shown in Ref. \cite{Barnett:1973b}, that the left hand side of \eqref{eq:rayleigh} is monotonically decreasing with velocity (below the limiting dislocation velocity), so that the Rayleigh velocity can be efficiently computed numerically.

\section{Examples}
\label{sec:examples}
%%%%%%%%%%%%%%%%%%%%%%%%%%%

\subsubsection*{The fcc slip systems}
%%%%%%%%%%%%%%%%%%%%%%

With the Burgers vector directions $\langle 110 \rangle$ and slip planes $\{111\}$, the pure screw dislocations fulfill the symmetry requirements for the limiting velocity $\vc^\txt{screw}$ to be given by \eqnref{eq:vcscrew} with the rotated elastic constants \cite{Blaschke:2020MD}
\begin{align}
c'_{44} & = \frac1{3}(c_{44}+2c')\,, &
c'_{45} & = \frac{\sqrt{2}}{3}(c_{44}-c')\,, &
c'_{55} & = \frac13(c'+2c_{44})
\,,
\end{align}
with $c'=(c_{11}-c_{12})/2$.
Pure edge dislocations, on the other hand do \emph{not} feature a reflection symmetry, but the determinantal  equation $\det(nn)=0$ is simple enough to determine in this case analytically that $\vc^\text{fcc,edge}=\sqrt{\min(c_{44},c')/\rho}$ coincides with the smallest shear wave speed of the fcc crystal.

\subsubsection*{The \{112\} slip planes in bcc crystals}
%%%%%%%%%%%%%%%%%%%%%%%%%

None of the pure screw dislocations in the 48 bcc slip systems with Burgers vector directions $\langle 111 \rangle$ fulfill the reflection symmetry requirement, so their limiting velocities are determined from \eqnref{eq:detnn}.
The only family of slip systems where pure edge dislocations feature the reflection symmetry are the $\{112\}$ slip planes.
They fall into the most general sub-category with $c'_{26}\neq0$ as discussed in the first paragraph of Sec. \ref{sec:edgereflection}, and are determined numerically from the solutions to the quartic equation \eqref{eq:quarticeq_vc}.

\subsubsection*{Basal slip in hcp crystals}
%%%%%%%%%%%%%%%%%%%%%%%%%%

With Burgers vector directions $\langle \bar2110 \rangle$ and slip planes $\{0001\}$, the pure screw dislocations fulfill the symmetry requirements for the limiting velocity $\vc^\txt{screw}$ to be given by \eqnref{eq:vcscrew} with the rotated elastic constants
\begin{align}
c'_{44} & = c_{44}\,, &
c'_{45} & = 0\,, &
c'_{55} & = (c_{11}-c_{12})/2
\,.
\end{align}
The limiting velocity for pure edge dislocations is given by either \eqnref{eq:vcedge_special1} with $c'_{66}=c_{44}$ or by \eqnref{eq:vcedge_special2}, and which of these two solutions is the correct one depends on the numerical values of the elastic constants;
this needs to be determined on a case by case basis by checking if $q$ is greater or smaller than zero.
For example, one may check that the basal slip system of Zn falls into the second category ($q<0$), whereas Mg and Ti fall into the first ($q\ge0$).

\subsubsection*{Prismatic slip in hcp crystals}
%%%%%%%%%%%%%%%%%%%%%%%%%%

With Burgers vector directions $\langle \bar2110 \rangle$ and slip planes $\{\bar1010\}$, the pure screw dislocations fulfill the symmetry requirements for the limiting velocity $\vc^\txt{screw}$ to be given by \eqnref{eq:vcscrew} with the rotated elastic constants
\begin{align}
	c'_{44} & = (c_{11}-c_{12})/2\,, &
	c'_{45} & = 0\,, &
	c'_{55} & = c_{44}
	\,.
\end{align}
The limiting velocity for pure edge dislocations is given by either \eqnref{eq:vcedge_special1} with $c'_{66}=(c_{11}-c_{12})/2$ or by \eqnref{eq:vcedge_special2} in principle, though all metals checked by the present author (Cd, Mg, Ti, Zn, and Zr) fall into the former category.

\subsubsection*{Pyramidal slip in hcp crystals}
%%%%%%%%%%%%%%%%%%%%%%%%%%

With Burgers vector directions $\langle \bar2110 \rangle$ and slip planes $\{\bar1011\}$, the pure screw dislocations fulfill the symmetry requirements for the limiting velocity $\vc^\txt{screw}$ to be given by \eqnref{eq:vcscrew} with the rotated elastic constants
\begin{align}
	c'_{44} & = \frac{c^2c'+\frac34a^2c_{44}}{\frac34a^2+c^2}\,, &
	c'_{45} & = \frac{\frac{\sqrt{3}}{2}ac(c'-c_{44})}{\frac34a^2+c^2}\,, &
	c'_{55} & = \frac{c^2c_{44}+\frac34a^2c'}{\frac34a^2+c^2}
	\,,
\end{align}
with $c'=(c_{11}-c_{12})$ and where $a$ and $c$ denote the two lattice constants within and perpendicular to the basal plane.
Thus,
\begin{align}
\vc^\txt{pyr,screw} &= \sqrt{\frac{c_{44}c'\left(\frac34a^2+c^2\right)}{\rho\left(\frac34a^2c_{44}+c^2c'\right)}}
\,.
\end{align}
The limiting velocity for pure edge dislocations does not feature a reflection symmetry in this case and is hence determined numerically from \eqnref{eq:detnn}.

\subsubsection*{Dislocations of mixed character}
%%%%%%%%%%%%%%%%%%%%%%%%%%

The special cases outlined in the preceding examples entailed only pure screw or pure edge dislocations.
Dislocations of arbitrary mixed character always need to be calculated along the lines of Section \ref{sec:generalcase}.

\section{Conclusion}
%%%%%%%%%%%%%%%%%%%%%%%%

In this short paper, we have reviewed how to compute the limiting velocities of pure screw and edge as well as mixed dislocations in arbitrary slip systems and crystal geometries.
We emphasized once more that these do not necessarily coincide with the lowest share wave speed of sound waves traveling in the dislocation glide direction, contrary to common lore.
In Sec. \ref{sec:examples}, we presented a number of simple examples for typical slip systems in cubic and hexagonal crystals.
A reference implementation of how to compute limiting velocities in the most general case is given within the open source code PyDislocDyn \cite{pydislocdyn} developed by the present  author.

\subsection*{Acknowledgements}
%%%%%%%%%%%%%%%%%%%%%%%%%%%%%%%%%%%%%%%%%%
\noindent
The author is grateful for the support of the Materials project within the Advanced Simulation and Computing, Physics and Engineering Models Program of the U.S. Department of Energy under contract 89233218CNA000001.

%%%%%%%%%%%%%%%%%%%%%%%%%%%%%%%%%%%%%%%%%%%
\bibliographystyle{utphys-custom}
\bibliography{dislocations}

\end{document}